# Teaching About Nature's Nuclear Reactors

J. Marvin Herndon

Transdyne Corporation
San Diego, California 92131 USA

July 10, 2005

## Abstract

Naturally occurring nuclear reactors existed in uranium deposits on Earth long before Enrico Fermi built the first man-made nuclear reactor beneath Staggs Field in 1942. In the story of their discovery, there are important lessons to be learned about scientific inquiry and scientific discovery. Now, there is evidence to suggest that the Earth's magnetic field and Jupiter's atmospheric turbulence are driven by planetary-scale nuclear reactors. The subject of planetocentric nuclear fission reactors can be a jumping off point for stimulating classroom discussions about the nature and implications of planetary energy sources and about the geomagnetic field. But more importantly, the subject can help to bring into focus the importance of discussing, debating, and challenging current thinking in a variety of areas.

Communications: mherndon@san.rr.com    http://NuclearPlanet.com

## Introduction

Nuclear reactor – the mention of those two words might bring to mind names like Chernobyl or Three Mile Island or perhaps conjure images of complex mega-machines whose control rooms have more instrumentation than the cockpit of a 747. But some nuclear reactors, the naturally occurring ones, are also a part of nature. Indeed, we may owe our well-being, if not our very existence, to a nuclear reactor at the center of the Earth.

Why should science teachers want to know about natural nuclear reactors? First, natural nuclear reactors are very much a part of our world and a subject at the forefront of scientific investigations. And, the subject matter falls well within NSE Standards for grades 9-12 and is appropriate for college classes as well. But in the story of their discovery, there are important lessons to be learned about scientific inquiry and scientific discovery. The subject of planetocentric nuclear fission reactors can be a jumping off point for stimulating classroom discussions about the nature and implications of planetary energy sources and about the geomagnetic field. But more importantly, the subject can help to bring into focus the importance of discussing, debating, and challenging current thinking in a variety of areas, instead of viewing science simply as an assemblage of facts (some of which may not even be facts). Science is about understanding, insight, and ideas. It is about reasoning.

**Discovery**

In the waning months of 1938, as the clouds of war began to shroud Europe, Otto Hahn and Fritz Strassmann bombarded a sample of uranium with neutrons. Afterward, they analyzed the sample to determine what elements might have been produced. On the basis of what was known at the time, Hahn and Strassmann expected perhaps to find different elements that were close to uranium on the periodic table, elements produced by neutron bombardment that differed from uranium at most by one or two atomic numbers. Imagine their surprise when they found instead the element barium, an element with nearly one-half of the atomic number and mass of uranium. As they reported early in 1939 in the German science journal, *Naturwissenschaften*, the neutrons had apparently split the nucleus of the uranium atom into two pieces, roughly equal in mass. They had split the atom, a process that subsequently became known as nuclear fission.

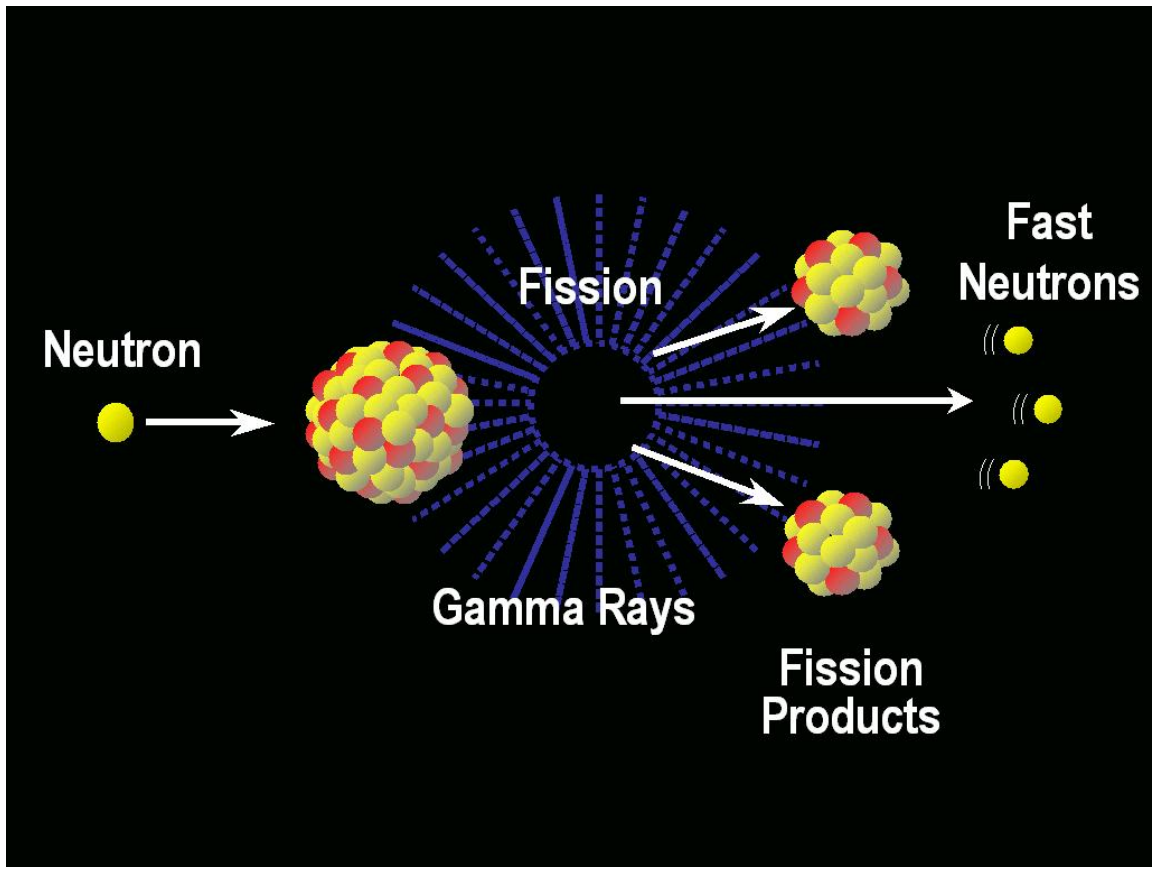

Later in 1939, Siegfried Flügge, writing in the same journal, speculated on the possibility that uranium chain reactions, capable of releasing enormous amounts of energy, might have taken place in the past in uranium deposits. But many years passed before the idea of naturally occurring nuclear fission reactors was taken seriously. To understand why, we need to look briefly at the process of nuclear fission.

Natural uranium at the present time contains about 136 times as many atoms of the essentially non-fissionable $^{238}$U (pronounced U-238) as it does the readily-fissionable $^{235}$U. Unlike $^{238}$U, $^{235}$U can easily fission, split into two parts, when its nucleus is hit by a neutron. The operation of a nuclear reactor depends upon maintaining a chain reaction.

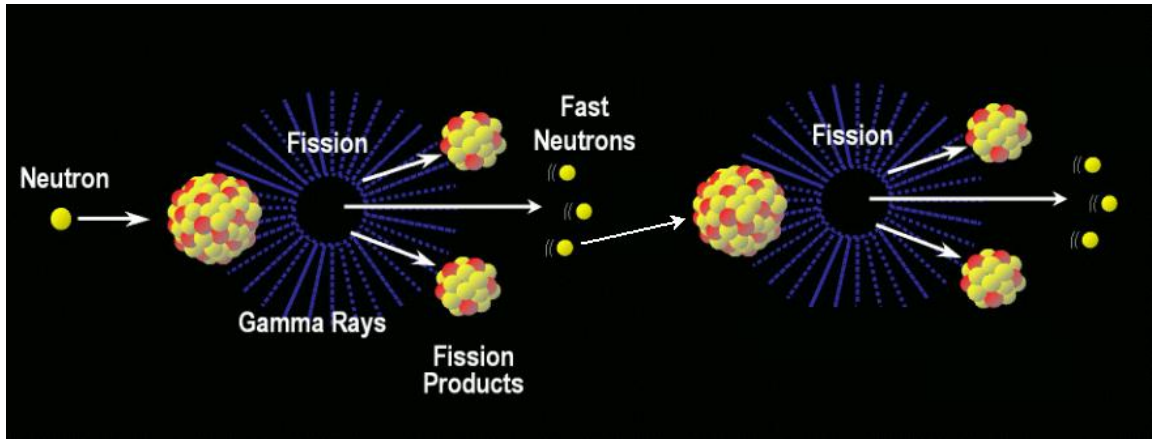

This is how it works: When a neutron enters a $^{235}$U nucleus causing it to fission, the nucleus usually splits into two "fission fragments" plus – and this important – typically two or three neutrons. If circumstances are right, one of those neutrons can cause another $^{235}$U nucleus to split, releasing two or three more neutrons, which in turn can cause yet another $^{235}$U nucleus to split, and so forth as the chain reaction proceeds. Each time that a $^{235}$U nucleus is split, energy is released.

So what are the circumstances needed for a nuclear reactor to be able to maintain a chain reaction? In the simplified picture, there are basically two considerations. First, the uranium has to be compact and thick enough so that not too many neutrons escape. Second, the amount of neutrons swallowed up by nuclei other than the fissionable $^{235}$U must be limited. $^{238}$U swallows up neutrons and its present-day high relative abundance is a real problem for using making a nuclear reactor that uses natural uranium as a fuel. These are the main points, but other considerations can be involved, for example, controlling neutron speed.

Enrico Fermi formulated nuclear reactor theory and in 1942 designed and built the world's first man-made nuclear reactor. It wasn't called a nuclear reactor then. It was called a pile, because that is essentially what it was, a stack of carbon blocks that were carefully embedded with pieces of uranium, a very imaginative solution to what seemed an almost impossible problem, making a nuclear reactor using natural uranium with its high proportion of neutron-swallowing $^{238}$U. It is no wonder that the idea of natural nuclear reactors was ignored for years! So, then no natural reactors? Think again! That is what Paul K. Kuroda did in 1956.

Both $^{238}$U and $^{235}$U are radioactive, *i.e.*, they change over time to a different element, but *not* at the same rate. $^{238}$U decays much more slowly that $^{235}$U. Consequently, Kuroda reasoned, in the past the relative proportion of $^{235}$U was greater than at present. In 1956, using Fermi's nuclear reactor theory, Kuroda showed that neutron chain reactions could have occurred in uranium deposits 2 billion years ago and earlier. Much later, Kuroda told me that the idea was so unpopular that he was only able to publish the paper because one journal at the time would publish short papers *without* peer review.

Fast forward 16 years to 1972 when I was a graduate student. One day Marvin W. Rowe, my thesis advisor, rushed into the lab to tell me that his former thesis advisor, Paul K. Kuroda, had just learned that French scientists had discovered the intact remains of a natural nuclear reactor in a uranium mine at Oklo, in the Republic of Gabon in Western Africa. The reactor had functioned 2 billion years ago just as Kuroda had predicted. Later, other fossil reactors were discovered in the region. I remember thinking at the time that the discovery must have huge implications, but there were just too many pieces missing from the puzzle to progress further; it was like looking out into a very, very dense fog.

Over the next 20 years, without consciously realizing it, I began to fill in the pieces. For example, in the 1970s and on into the 1980s, I realized that discoveries made in the 1960s admitted a different possibility for the composition of the Earth's inner core. I subsequently showed that the compositions of the parts of the deep interior of the Earth are more oxygen-deficient than previously thought. One important consequence is that large amounts of uranium would be expected to exist within the Earth's core, instead of residing exclusively in the mantle and crust.

At one time scientists thought that planets do not produce energy, except for tiny amounts from the decay of a few radioactive elements; planets just re-radiate energy from the sun. Then, in the late 1960s, astronomers discovered that Jupiter radiates about twice as much energy as it receives from the sun. The same is true for Saturn and Neptune. For two decades planetary scientists thought that they had considered and eliminated all possible planetary-scale energy sources, declaring "by default" that Jupiter's internally-generated energy was left over from planetary formation some 4.5 billion years ago. To me in 1990 that explanation did not make sense; Jupiter is 98% a mixture of hydrogen and helium, both of which transfer heat quite efficiently.

Information from studies of the natural reactors at Oklo indicates that a planetocentric nuclear reactor, initiated 2 billion years ago or earlier, can under appropriate conditions continue functioning even into the present by "breeding", producing additional fissionable fuel from $^{238}$U. Applying Fermi's nuclear reactor theory to the giant planets, as Kuroda had done to terrestrial uranium deposits, I demonstrated the feasibility of planetary-scale nuclear reactors as the internal energy sources for Jupiter, Saturn, and Neptune and published the concept in *Naturwissenschaften* in 1992. Significantly, those are the same three giant planets that display atmospheric turbulence, presumably driven by their internally-produced energy.

For more than a century, since Karl Friedrich Gauss, we have known that the seat of the geomagnetic field lies at or near the center of the Earth. We also know that there is an energy source residing there that continuously supplies energy to sustain the magnetic field; otherwise the field would soon collapse. It was only a small step to extend the nuclear reactor concept to the center of the Earth, which I published in the *Journal of Geomagnetism and Geoelectricity* in 1993 and in the *Proceedings of the Royal Society of London* in 1994. Unlike other planetary-scale energy sources, the energy output of a nuclear reactor can be variable, possibly even shutting down and re-starting. I have suggested that the polarity reversals of Earth's geomagnetic field might in some way be traceable to such variable georeactor energy production. Interestingly, as I learned much later after his death in 2001, Paul K. Kuroda was one of the reviewers of my 1993 paper.

For the past three decades, scientists and engineers at Oak Ridge National Laboratory have developed sophisticated computer programs to simulate numerically the operation of different types of nuclear reactors. Georeactor simulations conducted with those programs have not only verified all of my previous work on the subject, but have in addition led to new, strong evidence for the existence of a nuclear reactor at the center of the Earth. Once in every 10,000 fission events, the nucleus splits into three pieces, one of which is very low in mass. The Oak Ridge results have shown that helium, both $^3$He and $^4$He, will be produced by the georeactor in just the ratios observed in helium escaping from the Earth which was discovered in the late 1960s. Previously, for three decades scientists had not known of a deep-Earth mechanism for the production of $^3$He, so the $^3$He was assumed to be primordial helium, trapped since the time planet Earth formed (see http://NuclearPlanet.com/helium.htm). See how one new bit of understanding can lead to another. Now scientists throughout the world are looking into the possibility of detecting anti-neutrinos produced by the georeactor (for example, see http://arXiv.org/hep-ph/0401221).

## Thoughts for the Science Teacher

For science teachers, there are a number of lessons to be learned, perhaps the most important of which relates to the importance of discussing, debating and challenging scientific ideas. But what about the students? What is the most important message that a student might carry away? Perhaps it is this: The subject of planetary nuclear reactors is an example of how a single mind, without institutional support, can advance the frontier of science in a major way. This is a triumph of individual understanding, insight and ideas and is an example to science students everywhere of just what is possible.

The following notes relate to some principles, procedures, and practices of ethical science, some of which seem to have been forgotten or never learned by many science practitioners. These should become part of a young scientist's early training. Science teachers may find these useful for initiating discussion and debate, especially about the responsibilities of being a scientist.

1.) Hahn and Strassmanm made their discovery of nuclear fission as a consequence performing an experiment, (carefully and objectively) analyzing their data, and finding an unexpected result. Note that they did not begin by hypothesizing nuclear fission.
2.) Ideas and speculation can be important elements of scientific discussion and communication, but they should always be clearly labeled as such.
3.) Scientific issues should be considered from all possible angles. This sometimes requires very hard thinking and/or "out of the box", imaginative thinking.
4.) Popularity or "consensus" is *not* the way to assess scientific correctness. Science is a logical process, not a democratic process.
5.) When an important contradiction arises in ethical science, the new idea should be discussed and debated. Experiments and/or theoretical considerations should be made. If the new idea is wrong, it should be refuted in the literature, preferably in the journal of original publication; otherwise, it should be acknowledged.
6.) It is important to keep in mind that when you think about $n$ things, there may be at least ($n$+1) things, the 1 being the one that you have not yet thought of. In the case of Jupiter, the 1 was nuclear fission.
7.) It is important to re-think old ideas in light of new information.

## Discuss, Debate, and Challenge

The purpose of science is to understand the true nature of the Earth and the Cosmos. The Earth and the Cosmos are as they are; scientists can't change that. The best that scientists can hope to do is to discover their true nature. Discussing, debating, challenging and sometimes replacing old, inadequate ideas are very much part of scientific progress.

Science teachers should be aware that there is often very serious opposition to new ideas. Galileo Galilei's remarkable discoveries of sunspots and the moons of Jupiter in his own time evoked responses reflecting the darker side of human nature (see http://NuclearPlanet.com/galileo quotes.htm).

An ethical scientist, like Galileo, is interested in the *true* nature of the Universe and all contained therein. But throughout history and continuing into own time, there are those who seem to be more interested in being the purveyors of a flawed vision of the world, than representing to people how the world actually is or making it possible for people to be exposed to different ideas. Recall that Galileo was subjected to house arrest and, worse, Giordano Bruno was burned at the stake in Rome. More recently, note that Alfred Wegner's ideas of continental drift were systematically ignored for fifty years before being modified and revitalized as plate tectonics, and indeed, plate tectonics is not the final theory of whole-Earth dynamics (see http://arXiv.org/astro-ph/0507001). In fact, many science teachers are completely unaware that the inner core of the Earth may not be partially crystallized iron, as was first thought by Francis Birch in 1940 or that discoveries made in the 1960s admitted a different possibility (see http://NuclearPlanet.com/timetable.htm ).

Science is an investigation, a pursuit of truth that requires step-by-step logical thinking and unwavering integrity - values and ethics that are best learned at a young age. Discussing, debating, and considering other possible explanations are valuable approaches for training tomorrow's scientists. Not only does it begin to instill in students the approval to question scientific results, but it also provides a framework for considering the responsibilities that are and should be a part of ethical science**.**

**Thanks:** Jaroslav Franta graciously made the fission schematic representations.